\documentclass[preprint]{aastex} 
 
\slugcomment{MNRAS: in press} 
 
\shorttitle{Planetary nebula Hu\,2-1} 
\shortauthors{Miranda et al.}

\begin{document} 


\title{Morphological and kinematic signatures of a binary central star 
in the planetary nebula Hu\,2-1} 

\newcommand{\ap}{\mbox{a.p.}}                   
\newcommand{\arcmper}{\mbox{$\rlap.{' }$}}      
\newcommand{\arcsper}{\mbox{$\rlap.{'' }$}}     
\newcommand{\rlas}{\mbox{$\rlap.{'' }$}}        
\newcommand{\cc}{\mbox{cm$^{-3}$}}              
\newcommand{\tco}{\mbox{$^{13}$CO}}                             
\newcommand{\cdo}{\mbox{C$^{18}$O}}                             
\newcommand{\dec}{\mbox{$\delta(1950)$}}        
\newcommand{\dgr}{\mbox{$^\circ$}}              
\newcommand{\dper}{\mbox{$\rlap.^{\circ}$}}     
\newcommand{\et}{et al.\ }                      
\newcommand{\gapprox}{\mbox{$_>\atop{^\sim}$}}  

\author{Luis F. Miranda}
\affil{Instituto de Astrof\'{\i}sica de Andaluc\'{\i}a, CSIC, Apdo. Correos 
3004, E-18080 Granada, Spain.}
\email{lfm@iaa.es}

\author{Jos\'e M. Torrelles}
\affil{Institut d'Estudis Espacials de Catalunya (IEEC/CSIC) and Instituto de 
Ciencias del Espacio (CSIC), Edifici Nexus, C/ Gran Capit\'a 2-4, E-08034 
Barcelona, Spain}
\email{torrelles@ieec.fcr.es}

\author{Mart\'{\i}n A. Guerrero}
\affil {Department of Astronomy, University of Illinois at Urbana-Champaign,  
1002 West Green Street, Urbana, IL 61801 USA}
\email{mar@astro.uiuc.edu}

\author{Roberto V\'azquez}
\affil {Instituto de Astronom\'{\i}a, UNAM, Apto. Postal 877, 22800 
Ensenada, B.C., Mexico}
\email{vazquez@astrosen.unam.mx}

\author{Yolanda G\'omez}
\affil {Instituto de Astronom\'{\i}a UNAM, Campus Morelia, A.P. 3-72 
(Xangari), c.p. 58089 Morelia, Michoac\'an, Mexico}
\email{gocy@astrosmo.unam.mx}


\received{\today}
\accepted{}


\clearpage

\begin{abstract}

We present H$\alpha$, [NII] and [OIII] ground-based and HST archive images, 
VLA--A 3.6\,cm continuum and H92$\alpha$ emission line data and high 
resolution long-slit [NII] spectra of the planetary nebula Hu\,2-1. A large 
number of structural components are identified in the nebula: an outer 
bipolar and an inner shell, two pairs of collimated bipolar 
structures at different directions, monopolar bow-shock-like structures and 
an extended equatorial structure within a halo. The formation of 
Hu\,2-1 appears dominated by anisotropic mass ejection during the late 
AGB stage of the progenitor and by variable, ``precessing'' collimated 
bipolar outflows during the proto-planetary nebula and/or early planetary 
nebula phases. Different observational results strongly support the existence 
of a binary central star in Hu\,2-1, among them: (1) the observed 
point-symmetry of the bipolar lobes and inner shell, and the 
departures from axial symmetry of the bipolar lobes, (2) the off-center 
position of the central star, (3) the detection of mass ejection towards 
the equatorial plane, and (4) the presence of ``precessing'' collimated 
velocity of the bipolar shell. We propose that this velocity difference is 
a direct evidence of orbital motion of the ejection source in a binary 
system. From a deduced orbital velocity of $\sim$ 10 km\,s$^{-1}$, a 
semi-major axis of $\sim$ 9--27 AU and period of $\sim$ 25--80 yr are 
obtained, assuming a reasonable range of masses. These parameters are used to 
analyze the formation of Hu\,2-1 within current scenarios of planetary nebula 
with binary central stars.

\end{abstract}


\section{Introduction}

Binary stars are emerging as a basic ingredient in the formation and 
evolution of many planetary nebulae (PNe). Binary scenarios, involving 
different 
kinds of systems, have been invoked to provide qualitative explanations for 
different observations including, among others, the common asphericity of the 
PN shells and the existence of collimated outflows in PNe (e.g. Morris 1987; 
Soker \& Livio 1994; Soker 1996; Mastrodemos \& Morris 1998, 1999). On the 
other hand, only a scarce number of wide (separation $\sim$ 10$^2$--10$^4$ 
AU, period $\sim$ 10$^3$--10$^6$ yr) and close (separation $\sim$ 1--35 
R$_{\odot}$, period $\sim$ 0.11--16 days) binary central stars are presently 
known (Ciardullo et al. 1999; Soker 1999; Bond 2000 and references therein), 
most likely because direct detection is severely hampered by the present 
instrumental resolution, the nature of the binary itself and/or of its 
(stellar or sub-stellar) companion, and even by the evolution of the 
system (i.e., the binary may not exist anymore, e.g., Soker 1996). 
Therefore, a reasonable alternative to infer the possible presence and 
properties of a binary central star is to search for the influence of binary 
evolution in the nebular properties (e.g. Soker 1998).

Hu\,2-1 (PN\,G051.4+09.6) is a young PN whose properties suggest the 
existence of a binary central star. The main shell presents 
a bipolar morphology consisting of a bright equatorial toroid and two faint 
bipolar lobes (Aaquist \& Kwok 1990; Kwok \& Aaquist 1993, hereafter KA93; 
Miranda 1995, hereafter M95). An external ring-like structure, characterized 
by relatively high [NII]/H$\alpha$ values, surrounds the bright toroid (M95). 
Two highly collimated, high-velocity bipolar knots have been detected 
along the main nebular axis (PA $\simeq$ 320$^{\circ}$) via 
high-resolution, long-slit spectroscopy (M95). These knots present the 
typical properties of collimated outflows in PNe (e.g., Miranda 1999), 
although they also exhibit large differences in radial velocity and 
velocity width. Another collimated structure was also identified near the 
major nebular axis (M95). It shares some properties with the bipolar knots  
(small velocity dispersion, small size, strong [NII] emission) but it does 
not share the general inclination of the bipolar shell and the bipolar knots. 
On the basis of these observations, it has been suggested that a mass 
transferring binary central star could be involved in the formation of 
Hu\,2-1 (M95). There are additional structural components in this 
complex PN. The 3.6\,cm continuum map by KA93 reveals a small elliptical 
structure embedded in the toroid, suggesting that Hu\,2-1 is a double shell 
PN, and a faint extended halo has been detected in long-slit spectra (M95). 
Moreover, irregular variations of 0.2--0.3 mag in UVB have been reported 
(Kostyakova 1992; Arkhipova et al. 1994). Acker et al. (1998) analyzed 
HIPPARCOS data and measured a proper motion for Hu\,2-1 of $\simeq$ 
15$\times$10$^{-3}$ arcsec\,yr$^{-1}$ towards PA $\simeq$ 157$^{\circ}$. In 
addition, Acker et al. obtained a very small, unrealistic distance for 
Hu\,2-1, which, according to the authors, could be due to influence of the 
nebula itself in the parallax measurement. We will therefore adopt here a 
distance of 2.35 kpc (see M95).

We have obtained different data on Hu\,2-1 in order to study in more detail 
the different nebular components, testing the binary scenario. In this paper 
we present the results obtained.

\section{Observations and Results}

\subsection{Optical imaging}

Direct images of Hu\,2-1 were obtained with the 2.56\,m Nordic Optical 
Telescope (NOT)\footnote{The Nordic Optical Telescope is operated on the 
Island of La Palma by NOTSA in the Spanish Observatorio del Roque
de los Muchachos of the Instituto de Astrof\'{\i}sica de Canarias.} at Roque 
de los Muchachos Observatory in 1996 August. The detector was a Tektronix 
24$\mu$m CCD with 1024$\times$1024 pixels. The scale in the focal plane 
is $\simeq$ 0$\rlas$176 pixel$^{-1}$. The filters were: H$\alpha$ 
($\lambda$$_0$ $\simeq$ 6563 {\AA}, FWHM $\simeq$ 9 {\AA}), [NII] 
($\lambda$$_0$ $\simeq$ 6584 {\AA}, FWHM $\simeq$ 9 {\AA}) and [OIII] 
($\lambda$$_0$ $\simeq$ 5007 {\AA}, FWHM $\simeq$ 30 {\AA}). Exposure time 
was 20\,s in H$\alpha$, 180\,s in [NII] and 60\,s in [OIII]. Seeing was 
$\simeq$ 0$\rlas$72 for the three images.

Figure 1 presents contour plots of the H$\alpha$, [OIII] and [NII] images 
deduced from the ground-based data. Although the resolution of the 
ground-based images is lower than those of the HST (described below), we 
present the ground-based images because they are useful to compare with the 
long-slit spectra (\S\,2.3) and to obtain a global view of the excitation 
conditions in the nebula (see below). By comparing the images with the spectra 
previously presented (M95), the different structures can be identified. These 
are labeled in the [NII] image. Structure A corresponds to the bright 
equatorial toroid. It appears elongated along the nebular minor axis 
(PA 50$^{\circ}$) in H$\alpha$ and [NII] but circular in [OIII]. The bipolar 
lobes correspond to structure B. C1--C2 are the compact knots along the 
nebular major axis. C3--C4 are two compact structures at PA $\simeq$ 
350$^{\circ}$, reported for the first time in this paper. Component D is 
revealed to be an elongated structure oriented almost perpendicular to the 
nebular major axis. The images confirm that no counterpart of D exists 
towards the NW (M95). The existence of the extended (size $\simeq$ 14$''$) 
halo is confirmed. Structures within the halo can be recognized near the 
North-South direction, almost coinciding with C3--C4, and along PA $\simeq$ 
80$^{\circ}$. In addition, a bright region within the halo is observed along 
the nebular minor axis up to $\simeq$ 6$''$ from the center.

In order to gain a better view of Hu\,2-1 we have retrieved the WFPC2 
H$\alpha$, [NII] and PC [OIII] images obtained with the {\it Hubble Space 
Telescope} from the HST archive (program ID: 6347, PI: K.J. Borkowski \& J.P. 
Harrington for the H$\alpha$ and [NII] images; program ID: 3603, PI: 
M. Bobrowski for the [OIII] image). Figure 2 shows grey scale maps of 
the [NII] image and Figure 3 presents a grey-scale map of the central 
nebular regions as deduced from the H$\alpha$ image. The HST images confirm 
the basic bipolar structure of the nebula deduced from long-slit 
observations (M95). The bright toroid resembles a cylinder. If we assume 
circular cross section, the axis of the toroid is tilted $\simeq$ 
37$^{\circ}$ with respect to the plane of the sky, in excellent agreement 
with the inclination angle deduced from long-slit spectra (M95). A small 
structure within the toroid is also observed and corresponds to the inner 
elliptical shell identified by KA93. This inner shell is extremely faint in 
[NII] but very bright in H$\alpha$ and [OIII] (not shown here). The HST 
images show that this shell presents open ends (see also below). The bipolar 
lobes also present point-symmetry with respect to the central star but they 
are not symmetric with respect to the main nebular axis. In particular, if we 
consider PAs 310$^{\circ}$ and 330$^{\circ}$, which are symmetric with 
respect to the main axis (PA 320$^{\circ}$), the lobes are more extended at 
PA 330$^{\circ}$, radius $\simeq$ 2$\rlas$2, than at PA 310$^{\circ}$, 
radius $\simeq$ 1$\rlas$9. Finally, an elongated faint structure, 
particularly bright in H$\alpha$, is observed at PA $\simeq$ 80$^{\circ}$ 
and corresponds to that identified in the ground-based images. 

The central star is clearly observed in the HST images (Figs.2 and 3). As a 
remarkable result, we find that its position does not coincide with the 
center of the toroid but it is slightly displaced towards the SW along the 
nebular minor axis. A displacement of $\simeq$ 0$\rlas$05 (120 AU) is 
obtained from the [NII] and H$\alpha$ images with respect to the center 
of the toroid, and from the H$\alpha$ and [OIII] images with respect to the 
center of the inner shell.

The high resolution of the HST images allows us to study the structure of 
C1--C4 and D in detail. C1 and C3 consist each of two small and close knots. 
In the case of C1, the long-slit spectra (M95, see his Fig.4) show two knots 
at different radial velocity and separated by $\simeq$ 0$\rlas$9, in 
agreement with the HST [NII] image. In C2 no evidences for more than a knot 
are found, taking into account that part of the emission near the position 
of C2 corresponds to structure D (see M95). In the case of C4 a bright knot 
can be distinguished near the SE lobe and a fainter one at $\simeq$ 4$''$ 
from the center, which could be related to structure D. D appears as several 
bow-shock-like structures at different orientations. 

Figure 4 presents image ratios deduced from the ground-based images, which 
have been used to study the global variations of the excitation conditions in 
Hu\,2-1. A wealth of structure is observed in these images. The innermost 
nebular regions are of relatively high excitation whereas the edges of the 
bipolar lobes and the toroid present a relatively low excitation. 
Point-symmetry in the lobes can also be recognized in the excitation 
conditions. Low excitation is observed in C1 to C4 and D. The variation of 
the excitation along the nebular minor axis is particularly interesting. 
Close to the edges of the toroid (along PA 50$^{\circ}$), we observe a 
relatively low excitation zone. The excitation increases outwards and two 
high-excitation maxima are observed at $\simeq$ 2$\rlas$5 from the center. 
These maxima are very prominent in [OIII]/[NII] and seem to form part of a 
distinct high-excitation circular region which is better observed in 
[OIII]/H$\alpha$. Beyond this high-excitation region, H$\alpha$ dominates 
in a band along the equatorial plane up to $\simeq$ 6$''$ from the center. 
The ratio maps also show a region dominated by H$\alpha$ at 
PA $\simeq$ 80$^{\circ}$ up to $\simeq$ 4$''$ from the center (see above).

\subsection{Radio observations}

The observations were carried out with the Very Large Array (VLA) of the 
National Radio Astronomy Observatory (NRAO)\footnote{The NRAO is a facility 
of the National Science Foundation operated under a cooperative agreement with 
Associated Universities, Inc.} during 1995 July with the VLA in its A 
configuration. The data were obtained in the spectral line mode in order to 
study the H92$\alpha$ recombination line ($\nu$ = 8309.383 MHz [3.6\,cm]). We 
used a bandwidth of 6.25 MHz with two circular polarizations centered at the 
LSR velocity of 33.3 km\,s$^{-1}$ and 63 channels of 97.7 kHz ($\simeq$ 3.5 
km\,s$^{-1}$) wide each plus a continuum channel (channel 0) which
contains the central 75\% of the total bandwidth. Integration time was 
$\simeq$ 2.5 hours with $\simeq$ 20\% of this time spent in calibration. 
3C\,286 and 1923+210 were used as flux and phase calibrators, respectively. 
Calibration and image processing were carried out with the Astronomical Image
Processing System (AIPS) of the NRAO. Continuum emission at 3.6\,cm and 
H92$\alpha$ emission were detected from Hu\,2-1. We show in Figure 5 
a grey-scale map of the 3.6\,cm continuum emission obtained with uniform 
weighting of the (u,v) data (beam = 0$\rlas$20$\times$0$\rlas$19, 
PA = $-$14$^{\circ}$). In Figure 6, we show contours of the H92$\alpha$ 
emission obtained by adding all channels in which line emission is 
detected (from 19 to 51 km\,s$^{-1}$), superposed on a 3.6\,cm continuum map 
made with natural weighting and a Gaussian taper of 500 k$\lambda$ 
(beam = 0$\rlas$42$\times$0$\rlas$40, PA = 60$^{\circ}$). The continuum flux 
density at 3.6\,cm is $\simeq$ 110 mJy, compatible with the value obtained 
by KA93. The H92$\alpha$ flux density is $\simeq$ 3.9 mJy. 

The 3.6\,cm continuum map (Fig.5) shows the innermost regions of Hu\,2-1 in 
considerable detail. The bright regions separated $\simeq$ 1$''$ at 
PA $\simeq$ 50$^{\circ}$ correspond to the outer edges of the toroid. 
The inner shell, identified by KA93, can be distinguished with a size of 
$\simeq$ 0$\rlas$35 and with the major axis near PA 320$^{\circ}$. In our 
map, it shows open ends and a clear point-symmetric brightness distribution. 
Detection of H92$\alpha$ emission is restricted to a region of 
$\simeq$ 0$\rlas$5 in radius apparently coinciding with the inner shell 
(Fig.6). We note that given the intensity of the continuum emission in the 
outer edges of the toroid, one would expect to detect H92$\alpha$ emission 
from these positions as well. However, the continuum emission towards the 
edges of the toroid is probably optically thick, preventing the detection of 
H92$\alpha$ emission. 

Figure 7 presents the H92$\alpha$ line profile. The LSR central velocity and 
the line width (FWHM) of the H92$\alpha$ emission have been determined by 
means of a single-gaussian fit and are listed in Table 1. Given the reduced 
S/N of the profile, the deduced systemic velocity (V$_{LSR}$ $\simeq$ 37 
$\pm$ 2 km\,s$^{-1}$) is compatible with that deduced from optical emission 
lines (see below). Because of the low S/N in the individual channels, a 
detailed study of the kinematics in H92$\alpha$ is not possible. In order to 
increase the S/N, we have added the blueshifted channels and the redshifted 
channels with respect to the systemic velocity as deduced from the optical 
lines. The result is presented in Figure 8. Blueshifted emission appears 
slightly displaced towards the N-NW, with respect to the emission center 
in the continuum image, whereas redshifted emission is displaced 
towards the S-SE.

Estimates for the electron density and electron temperature were obtained 
from the continuum and line emission following the formulation by Mezger \& 
Henderson (1967), assuming the gas is in LTE and from the observed 
line-to-continuum ratio T$_C$/T$_L$ of $\sim$ 12.5. The results are listed in 
Table 1. The values obtained for the electron density and temperature are 
similar to those obtained from optical emission lines (see M95 and references 
therein).

\subsection{Long-slit spectroscopy}
 
Long-slit spectra were obtained in 1998 August with IACUB\footnote{The IACUB 
uncrossed echelle spectrograph was built in a collaboration between the IAC 
and the Queen's University of Belfast} at the NOT. A filter was used to 
isolate the H$\alpha$ and [N\,{\sc ii}]6583 emission lines. Two spectra were 
obtained at PAs 320$^{\circ}$ and 351$^{\circ}$ with the slit centered on the 
object. Exposure time was 900\,s for each spectrum. A Th-Ar lamp was 
used for wavelength calibration. The spectral resolution (FWHM) is $\simeq$ 
8 km\,s$^{-1}$. Seeing was $\simeq$ 1.4$''$. The data were reduced 
using standard procedures within the IRAF\footnote{IRAF is distributed by 
the National Optical Astronomy Observatory, operated by the Association of 
Universities for Research in Astronomy, Inc. (AURA) under cooperative 
agreement with the National Science Foundation} and MIDAS packages. The error 
in absolute radial velocity is estimated to be $\simeq$ $\pm$ 1.2 
km\,s$^{-1}$. However, the internal accuracy in each long-slit spectrum is 
much better, $\simeq$ $\pm$ 0.15 km\,s$^{-1}$. Contour plots of the [NII]6583 
emission line, in a position--velocity representation, are shown in Figure 9. 
As compared with the spectra presented by M95, the new ones, although with 
poorer spatial resolution, have a higher spectral resolution and provide 
information about the bipolar knots C3--C4. 

Table 2 lists different parameters of the knots C1 to C4 (PA, radial velocity, 
angular position and spectral and angular widths) obtained from the spectra. 
Radial velocities and angular positions refer to the centroid of the 
knots. The radial velocities are quoted with respect to the systemic velocity 
of the main nebular shell [V$_{sys}$(shell)] (hereafter considered as the 
toroid and bipolar lobes) for which we deduce V$_{LSR}$ = 34.5 $\pm$ 1.2 
km\,s$^{-1}$ (V$_{\odot}$ = 15.1 km\,s$^{-1}$) in agreement with previous 
determinations (Schneider et al. 1983; M95; Durand et al. 1998). Spatial 
positions have been determined with respect to the central star, represented 
by the position of the maximum of the stellar continuum in the 
two--dimensional frames, with a precision of $\simeq$ $\pm$ 0$\rlas$1. 
The results for C1--C2 are compatible with those obtained by M95, the 
small differences being due to the fact that in Table 2 the parameters 
refer to the centroid of the knots, whereas M95 considered the peak 
intensity. In addition, the new spectra show that C3--C4 constitutes a system 
of highly collimated bipolar knots.

The following results are noticed from Table 2. The radial velocity of the 
NW knots C1 and C3 is systematically higher than that of their corresponding 
SE counterparts C2 and C4. In particular, the radial velocity of C1 is 11 
km\,s$^{-1}$ higher than that of C2 and the radial velocity of C3 is 
16 km\,s$^{-1}$ higher than that of C4. The radial velocity decreases with 
the distance in C1 and C3. In fact, the position--velocity maps (Fig.8) 
suggest that two knots at different velocities are present in C3. Within 
C2 and C4 the radial velocity is constant (see also M95). C1--C2 are located 
at larger distance from the central star than C3--C4. However, differences 
in the distance for the knots in a particular pair do not seem to be 
significant. C1 and C3 present a higher velocity width than C2 and C4. This 
can be explained taking into account that C1 and C3 consist of two small 
knots at different velocity (see above) and, therefore, the velocity width 
results from the superposition of two knots which are not spatially resolved 
in our spectra.

The results for the toroid and bipolar lobes are compatible with those 
obtained by M95. In the case of D we confirm the previous results, in 
particular, its very low radial velocity (between $-$2 and $+$2 km\,s$^{-1}$, 
see M95 for details).

\section{Discussion}

In the following we will discuss possible interpretations for the data 
presented above, testing the scenario of a binary system at the center of 
Hu\,2-1.

\subsection{Shaping of the inner and bipolar shells}

The data suggest that the inner shell could be (1) an elliptical, independent 
shell (see KA93), (2) the inner ``walls'' of the toroid or (3) a second, 
inner toroid concentric with the outer one. The kinematics of this shell can 
be used to decide between these possible interpretations. As already 
mentioned, the H92$\alpha$ emission seems to be related to the inner shell. 
If this shell represents the inner walls of the toroid or a second toroid 
concentric with the outer one, it would be expected to detect the 
H92$\alpha$ emission blueshifted (redshifted) towards the SE (NW) part of 
the toroid, according to its inclination (M95). However, the opposite 
is found (\S 2.2, Fig.8) as expected if the H92$\alpha$ emission would trace 
an expanding ellipsoidal shell with the NW (SE) part blueshifted (redshifted) 
as the bipolar lobes or a toroid with a different inclination as the outer 
one. Nevertheless, these interpretations do not explain the point-symmetry of 
the inner shell, which suggests the existence of a non-spherical wind 
interior to the inner shell, at scales $\leq$ 0$\rlas$35, which is not 
oriented along the major nebular axis and impacts the inner shell at 
certain angle. It is possible that the H92$\alpha$ emission could be tracing 
a bipolar outflow interior to the inner shell. The foreseen upgrading of the 
VLA will certainly allow us to study with great sensitivity and high 
resolution the innermost regions of the nebula.

With regards to the bipolar shell, the equatorial toroid expands at $\simeq$ 
15 km\,s$^{-1}$ in [NII] (M95) and the HST [NII] image indicates a size of 
$\simeq$ 1$\rlas$1, so that its kinematic age is $\simeq$ 410 yr. The bipolar 
lobes expand at $\simeq$ 30 km\,s$^{-1}$ (M95) and the HST images indicate a 
maximum radius of $\simeq$ 2$\rlas$2 at PA 330$^{\circ}$. The bipolar axis is 
tilted $\simeq$ 37$^{\circ}$ with respect to the plane of the sky. With these 
numbers, the kinematic age of the bipolar lobes results to be $\simeq$ 
1000 yr, much larger than the age of the toroid. This difference can 
be explained if the kinematics of the lobes is more complex than previously 
assumed (M95). In fact, the maximum size of the lobes is observed along 
PA 330$^{\circ}$ where the largest departures from axial symmetry are 
observed. It is plausible that these regions are expanding at a higher 
velocity than $\simeq$ 30 km\,s$^{-1}$ and/or are tilted at an angle 
different from 37$^{\circ}$. Therefore, the kinematic age of the toroid 
probably is more representative for the age of the bipolar shell.

In any case, the shape of the bipolar lobes points out to a bipolar wind 
interior to the main shell. In an idealized model of interacting stellar 
winds (see Zhang \& Kwok 1998 and references therein), the lobes in a 
bipolar PN should be symmetric with respect to the bipolar axis. This is not 
the case for the lobes in Hu\,2-1, which show clear departures from 
axis-symmetry. As in the case of the inner shell, we propose the 
existence of an interior bipolar wind which is not oriented along the main 
nebular axis, but impacts the lobes at certain angle, so that the lobes are 
inflated in an asymmetric (with respect to the main nebular axis) but 
point-symmetric (with respect to the central star) manner. Interaction of 
collimated outflows with shells has been suggested to occur in other PNe 
(Sahai \& Trauger 1998; Miranda et al. 1999; V\'azquez et al. 1999, 2000). We 
note that in the case of Hu\,2-1, however, knots C1 to C4 do not seem to be 
involved in shaping the bipolar lobes, because evidence for interaction 
between the knots and lobes cannot be recognized in our data. If interaction 
would have existed, one would expect strong deformations of the lobes at the 
PAs where the knots are detected (see, e.g., Miranda et al. 1999), which is 
not observed. The origin of the interior bipolar wind is difficult to 
elucidate from the observed geometry alone. It could be either that the fast 
stellar wind from the central star is anisotropic in origin or that two winds 
are present: an isotropic fast stellar wind from the central star and an 
intrinsically  bipolar wind from a companion. 

\subsection{The structure of the equatorial plane}

The image ratio maps show an equatorial zone within the halo, which is 
characterized by peculiar line ratios, particularly those involving 
H$\alpha$. This suggests the existence of a flat equatorial region with 
physical conditions different from those in other parts of halo. A 
plausible interpretation for this equatorial region is that it is related 
to, and traces anisotropic mass ejection during the AGB phase of the 
Hu\,2-1 progenitor. In fact, for an expansion velocity of 10 km\,s$^{-1}$, 
the kinematic age of the flat region is $\simeq$ 7000 yr, much larger 
than that of the main shell. Taking into account that the transition time 
from AGB to PN is $\simeq$ 2000--3000 yr and the evolutionary time in the 
AGB is $\simeq$ 10$^6$ yr, the formation of this region has occurred in the 
very late stages of AGB evolution. The distinct high-excitation region 
within the halo is younger, $\simeq$ 2800 yr, also assuming a velocity of 
10 km\,s$^{-1}$. This region could represent the final ejection in the AGB 
before the object entered in its proto-PN phase. The high-excitation in this 
region could be related to hardening of the radiation as observed in 
high-excitation haloes of PNe (Guerrero \& Manchado 1999). 

Different mechanisms have been suggested to produce anisotropic mass loss 
during the AGB, including stellar rotation, magnetic fields, non-radial 
pulsations and binariety (Soker 1996 and references therein; 
Garc\'{\i}a-Segura et al. 1999; Mastrodemos \& Morris 1999). In many cases, 
the presence of a companion is required either to spin up the envelope of 
the AGB star or to deflect the mass towards the equatorial plane (see, e.g., 
Soker 1996). However, Garc\'{\i}a-Segura et al. (1999) have recently modelled 
single stars rotating at near-critical rotation rates at the end of AGB. These 
models also produce ejection towards the equatorial plane and, when 
combined with an interacting winds scenario and magnetic fields, they are 
capable to account for the large variety of shapes observed in PNe 
(Garc\'{\i}a-Segura et al. 1999). In the case of Hu\,2-1, the simple 
detection of an equatorial ring-like structure ejected during the end of 
AGB does not allow us to discriminate between single star or binary scenarios.

\subsection{The off-center position of the central star}

Off-center central stars are observed in many PNe (see Soker 1999 and 
references therein). We compare Hu\,2-1 with MyCn\,18 (Sahai et al. 
1998), which are considered to be at a similar distance. In both cases, the 
central star is displaced along the nebular minor axis. In Hu\,2-1,  
the displacement is $\simeq$ 120 AU while larger displacements, between 
270 and 1150 AU, are found in MyCn\,18 with respect to different nebular 
structures (Sahai et al. 1998). A possibility to explain a displaced central 
star is relative proper motion of the central star and nebula. In Hu\,2-1 this 
possibility can be ruled out because the proper motion vector of the nebula 
is perpendicular to the direction of displacement (Acker et al. 1998). 
Another possibility to be considered is binariety of the central star. 
Soker (1994) and Soker et al. (1998) model the influence in the nebular 
morphology of wide and eccentric close binaries, respectively. In both 
cases, off-center central stars are predicted. Hu\,2-1 fits much better  
the eccentric scenario because (1) the displacement is observed 
along the minor nebular axis and (2) bipolar PNe could be related to 
eccentric binaries (see Soker et al. 1998). According to these authors, 
orbital separations of 7--80 AU and orbital periods of 15--500 yr are 
suggested for this kind of eccentric close binaries (see \S\,3.5).

\subsection{Collimated outflows and possible detection of orbital motion}

The new data have revealed the existence of multiple collimated knots in 
Hu\,2-1 at different directions. Whereas C1--C2 and C3--C4 represent 
the typical bipolar collimated outflows observed in PNe, the nature of D is 
less clear. The bow-shock appearance suggests that D could represent a series 
of collimated outflows. In this case, the outflows should have been monopolar 
and either they move almost perpendicular to the line of sight or its 
expansion velocity is very low. In any case, D does not share the general 
inclination of the nebula. We note that D is located towards the SE, 
coinciding with the direction of the proper motion of Hu\,2-1 (Acker et al. 
1998). Therefore, it is possible that the formation of D could be related to 
interaction of the nebula with the interstellar medium. 

The properties of the collimated outflows C1--C2 and C3--C4 suggest a 
precessing ejection source. These kind of outflows is much better accommodated 
by a binary scenario than into a single-star one (see below). Single-star 
models have been able to explain jets in PNe if stellar rotation and magnetic 
fields are considered (Garc\'{\i}a-Segura et al. 1999). This model 
produces jets along the main nebular axis. If precessing jets are to be 
explained within these models, a companion to the central star is required 
(Garc\'{\i}a-Segura 1997). Therefore, the mere presence of collimated 
outflows at different directions in Hu\,2-1 suggests a binary central star.

The most striking result concerning the bipolar outflows is the 
systematic difference between the radial velocities of the two knots in each 
bipolar pair. This could be attributed to differences in the ejection 
velocity (M95). In this case, systematic differences in position of the knots 
with respect to the central star would be expected, but they are not 
observed. Another possibility is deceleration of the knots 
by interaction with nebular material. If so, deceleration should have 
been more conspicuous in C2 and C4, the knots with lower radial velocity. 
However, evidence for deceleration exists in C1 and C3 which 
show an internal decrease of the radial velocity (see \S 2.3), but not in 
C2 and C4 in which the radial velocity is constant. We also note that a 
combination of different ejection velocity, different ejection angle and 
selective deceleration is highly improbable because of the very peculiar 
combination of these parameters necessary to explain the systematic 
differences in radial velocity and the symmetric location of the knots in 
each pair with respect to the central star. 

In order to explain these results, we consider a completely different and 
novel point of view, namely, that the differences in radial velocity arise 
only because we have measured them with respect to V$_{\rm sys}$(shell) which, 
in this case, is a misleading reference system for the collimated outflows. 
In other words, we suggest that the systemic velocity of the main nebular 
shell does not coincide with the systemic velocity of the collimated knots. 
Remarkably, this situation is what one would expect from a binary ejection 
source. The systemic velocity of the main nebula is related to material 
ejected during the last stages of the AGB phase averaged over a relatively 
large time span and over many orbits. In a first approximation, the average 
velocity of the ejection will be the same in all directions and the systemic 
velocity will not contain information about the orbital velocity or about a 
particular orbital position (but see below). The collimated knots, however, 
can be considered as ``instantaneous'' ejections, given its compactness. 
Their space velocities will contain two components: the component due 
to the own ejection velocity and the component due to the ``instantaneous'' 
orbital velocity at the time of ejection. The component due to the orbital 
velocity will be added with the same sign to the component due to its own 
ejection velocity in each knot of a bipolar pair. As a consequence of the 
additional orbital component of the velocity, the systemic velocity of the 
knots will be shifted with respect to V$_{\rm sys}$(shell). 

The data in Table 2 allow us to obtain the systemic velocity and the 
radial velocities of the knots with respect to their own systemic velocity. 
For C1--C2 we obtain V$_{\rm sys}$(C1--C2) $=$ 29 km\,s$^{-1}$ and a radial 
velocity of $\pm$ 53.5 km\,s$^{-1}$ with respect to V$_{\rm sys}$(C1--C2). 
For C3--C4, V$_{\rm sys}$(C3--C4) is 26.5 km\,s$^{-1}$ and the radial 
velocity is $\pm$ 58 km\,s$^{-1}$ with respect to V$_{\rm sys}$(C3--C4). The 
three systemic velocities V$_{\rm sys}$(C1--C2), V$_{\rm sys}$(C3--C4) and 
V$_{\rm sys}$(shell) are represented in Fig.9. Both V$_{\rm sys}$(C1--C2) and 
V$_{\rm sys}$(C3--C4) are blueshifted with respect to V$_{\rm sys}$(shell) 
by 5.5 and 8 km\,s$^{-1}$, respectively. These values are noticeable larger 
than the relative errors (\S2). In addition, V$_{\rm sys}$(C1--C2) and 
V$_{\rm sys}$(C3--C4) are almost identical to each other within the absolute 
errors. In the following, we will consider as radial velocity of the knots 
that referred to their own systemic velocity.
 
Estimates for the expansion velocity, distance to the central star and 
kinematic age of the collimated outflows can be obtained if the inclination 
angle is known. The location of C1--C2 along the major nebular axis suggests 
that C1--C2 move along that axis (M95). However, this must not be necessarily 
true. For instance, some parts of D also project along the major axis but 
they move in a completely different direction. In addition, we note that 
for inclinations of C1--C2 $\leq$ 40$^{\circ}$, the kinematic age of the 
knots is lower than that of the toroid. This would imply interaction of the 
collimated knots with the shell, which is not observed (see above). 
Therefore, we will assume a range of 40$^{\circ}$--60$^{\circ}$ for the 
inclination of C1--C2 and a constant velocity for the two pairs. We obtain a 
range of outflow velocities between 60--85 km\,s$^{-1}$ and a range of 
kinematic ages for C1--C2 between 520--1100 yr. The difference of kinematic 
ages between C1--C2 and C3--C4 is $\simeq$ 15--100 yr.

\subsection{Binary parameters and implications for jet formation in Hu\,2-1}

The available information can be used to constrain the characteristics of 
the binary central star in Hu\,2-1. From the simple model outlined above 
(\S 3.4), it can be easily demonstrated that 
\begin{equation}
V_{sys}(knots) - V_{sys}(shell) =  V_{orb} \times cos(\gamma) \times \cos(i)
\end{equation}
where V$_{\rm orb}$ is the orbital velocity at the time of ejection, 
$\gamma$ is the angle between the orbital velocity vector and the line of 
sight and $i$ is the inclination angle of the orbital plane with respect 
to the line of sight. For V$_{\rm sys}$(knots) $-$ V$_{\rm sys}$(shell) we 
obtain 5.5--8 km\,s$^{-1}$. In addition, it is reasonable to assume that the 
orbital plane coincides with the equatorial plane of the nebula, which is 
tilted $\simeq$ 37$^{\circ}$. With these assumptions, we obtain 
V$_{\rm orb}$$\times$cos($\gamma$) $\simeq$ 10 km\,s$^{-1}$, which is a lower 
limit to the orbital velocity. For masses M$_1$$+$M$_2$ in the range 
1--3\,M$_{\odot}$, we obtain upper limits for the orbital separation of 
$\simeq$ 9--27 AU (assuming circular orbit, see also below) and for the 
period of $\simeq$ 25--80 yr. It is worth noting that these orbital 
parameters are in the range of those expected from the displacement of the 
central star (\S 2.1). Therefore, the true orbital separation and period 
should not be much lower than the obtained upper limits, which also implies 
that the orbital velocity should not be much larger that the deduced value.

Two basic binary scenarios have been suggested to explain the generation of 
precessing collimated outflows in PNe. A scenario is that of a mass 
transferring binary (Morris 1981; Soker \& Livio 1994; Mastrodemos \& Morris 
1998). In this scenario, mass lost from the AGB star is partially captured by 
the secondary forming an accretion disk around it, from which collimated 
outflows are produced. Precession of the accretion disk will result in 
changes of the direction of the collimated outflows. A different scenario is 
considered by Soker \& Livio (1994) and Soker (1996), in which a disk is 
formed around the AGB nucleus by Roche lobe overflow and destruction of a 
stellar or substellar companion. In this case, the jets emanate from the 
AGB star. Radiation may induce self-warping of the disk so that the disk 
precesses originating precessing outflows (Livio \& Pringle 1996). In 
the first case, models by Mastrodemos \& Morris (1998, 1999) indicate that 
accretion onto the secondary is effective at large orbital separations 
($\simeq$ 24 AU). The second scenario requires orbital separations 
$\leq$ 2\,R$_{\odot}$ for the secondary being destroyed (Soker 1996; 
Reyes-Ruiz \& L\'opez 1999). The binary parameters deduced for Hu\,2-1 
clearly favor a mass transferring binary scenario with the collimated ejection 
generating from an accretion disk around the secondary. Furthermore, the 
ejection of material towards the equatorial plane is also compatible with 
a binary of these characteristics (Mastrodemos \& Morris 1999). 

In the previous calculations, we have assumed circular orbit. This is not 
necessarily true. In fact, the off-center position of the central 
star points out to an eccentric binary (\S\,3.3). In this case, the ejection 
velocity of the AGB wind may depend on the orbital position (Soker 1998; 
Mastrodemos \& Morris 1998) so that the measured systemic velocity of the 
nebula will contain some information about the orbital velocity. However, the 
displacement of the central star in Hu\,2-1 is small as compared, for 
instance, with that in MyCn\,18. In fact, the displacement amounts 
$\simeq$ 5--15\% in terms of the size of the outer edges of the toroid and 
inner shell, respectively, which is smaller than the displacement observed 
in MyCn\,18 ($\simeq$ 25\%). This suggests that the ellipticity is small 
and/or that there is a small dependence of mass loss with the orbital 
position. However, without regards of these considerations, the systemic 
velocities of the two pairs are very similar to each other. From this result 
and taking into account that the amplitude of the radial velocity curve could 
be $\sim$ 20 km\,s$^{-1}$ (see above), it can be concluded that collimated 
ejection occurs at a particular orbital phase and may be a ``periodic'' 
phenomenon. This suggests that mass accretion is enhanced at some points of 
the orbit, e.g., through passage by the periastron so that ``massive'' 
collimated outflows in the form of bright knots are generated. This 
conclusion is supported by the similarity between the difference of 
kinematic ages of the two collimated pairs (15--100 yr) and the orbital 
period (25--80 yr). At different orbital positions, after ``massive'' 
collimated ejection occurs, the bipolar outflow could still be active but 
with different properties (e.g., mass loss, velocity, direction, collimation 
angle) so that collimated knots are not generated. However, the bipolar 
outflow may still be capable to shape the bipolar lobes and the inner shell.

The orbital parameters deduced in Hu\,2-1 are remarkably similar to those 
of symbiotic stars (e.g. Seaquist \& Taylor 1990). Some symbiotic stars 
exhibit bipolar shells, equatorial disks and collimated outflows. In 
particular, we find remarkable similarities between Hu\,2-1, R\,Aqr and 
HM\,Sge. The orbital separation and period in these symbiotic are: 
18 AU and 44 yr in R\,Aqr and 25 AU and 90 yr in HM\,Sge (Hollis et al. 1997; 
Richards et al. 1999). Although these parameters are rough estimates, they 
compare very well to those of Hu\,2-1. Moreover, ``precessing'' collimated 
outflows are present in R\,Aqr and HM\,Sge as well as a bipolar shell and an 
equatorial ring-like structure (see Solf 1984; Solf \& Ulrich 1985; Hollis et 
al. 1990; Corradi et al. 1999). We also note that Solf \& Ulrich (1985) found 
in R\,Aqr a difference of 13 km\,s$^{-1}$ between the systemic velocity of the 
two bipolar shells, which was attributed to orbital motion. This suggestion 
is supported by the velocity semi-amplitude for the Mira in R\,Aqr 
(Hollis et al. 1997). The possible evolutionary relationship of symbiotic 
stars with PNe has already been pointed out (Corradi et al. 1999 and 
references therein). From the results presented in this paper, we suggest 
that Hu\,2-1 represents the result of the evolution of a symbiotic star 
similar to R\,Aqr and HM\,Sge, after the AGB star has evolved towards 
a central star of PNe.

\section{Conclusions}

We have carried out a multiwavelength observational study of the PN Hu\,2-1 
based on a set of optical/radio data obtained at high spectral and/or spatial 
resolution. These data have allowed us to identify many structural components 
in this complex PN. The main shell of Hu\,2-1 is bipolar and consists of a 
toroid and two point-symmetric lobes which also exhibit noticeable departures 
from axial symmetry. Inside the toroid, an inner point-symmetric shell is 
observed. Two bipolar pairs of low-excitation knots are identified at 
different directions and correspond to collimated outflows. Several 
monopolar, low-excitation bow-shocks-like structures are observed 
on one side of the nebula. We confirm the existence of an extended halo 
around the main nebula. This halo contains a distinct, extended equatorial 
disk-like region with peculiar excitation conditions.  

The early formation of Hu\,2-1 seems to be dominated by mass ejection towards 
the equatorial plane. This anisotropic mass loss process, resulting in the 
formation of a flat region within the halo, appears to have been active 
from the very late stages of AGB evolution up to the onset of the proto-PN 
phase. However, during the proto-PN and/or PN phase, variable, 
``precessing'' bipolar collimated outflows appear as the dominant formation 
mechanism of the nebula. This kind of outflows can be identified not only as 
two systems of highly collimated bipolar knots at different orientations but 
they could also be recognized in the point-symmetry of the bipolar lobes and 
inner shell and in the departure from axis-symmetry of the bipolar lobes. 
We also note that interaction with the interstellar medium could also be 
involved in the formation of the monopolar structures.

The whole set and the consistence of the different results provide strong 
support to conclude that a binary central star has been involved in the 
formation of Hu\,2-1. First-order estimates for the orbital parameters 
(separation $\sim$ 9--27 AU, period $\sim$ 25--80 yr) have been deduced 
from the observed kinematics and a simple binary model. These parameters 
suggest that the central star of Hu\,2-1 is a mass transferring binary in 
which collimated ejections are generated from an accretion disk around a 
companion. There are many similarities between Hu\,2-1 and symbiotic stars 
as, e.g., R\,Aqr and HM\,Sge, which lead us to suggest that Hu\,2-1 may be 
the descendent of this kind of binary stars.  

We suggest that a precise comparison of the systemic velocity of collimated 
outflows and main shell in other PNe may be a powerful method to detect the 
effects of orbital motion in a binary central star and/or to place strong 
constraints on the possible nature of the central star and on the scenario 
for the generation of collimated outflows.

\vspace{0.75cm}

\noindent {\bf Acknowledgments}   

We thank our referee for his/her comments which have been very useful to 
improve the presentation and interpretation of the data. We are very grateful 
to E. Alfaro, G. Anglada, A. Claret and E. P\'erez for useful comments and 
discussions. LFM and JMT are supported partially by DGESIC PB98-0670-C02 and 
Junta de Andaluc\'{\i}a (Spain). MAG is supported partially by DGESIC of the 
Spanish Ministerio de Educaci\'on y Cultura. YG acknowledges financial 
support from DGAPA-UNAM and CONACyT-Mexico. RV acknowledges financial support 
from CONACyT grant I32815E (Mexico). This work has been supported partially 
by the Programa de Cooperaci\'on Cient\'{\i}fica con Iberoam\'erica.

\newpage

\centerline{\Large\sc REFERENCES} 

\vspace{0.25cm}

\noindent Aaquist O.B., Kwok S., 1990, A\&AS, 84, 229

\noindent Acker A., Fresneau A., Pottasch S.R., Jasniewicz G., 1998, A\&A, 
337, 253

\noindent Arkhipova, V.P., Kostyakova E.B., Noskova R.I., 1994, PAZh, 20, 122

\noindent Bond H.E., 2000, in ASP Conf. Ser., vol.\,199, J.H. Kastner, N. 
Soker \& S. Rappaport (eds.), 115

\noindent Ciardullo R., Bond H.E., Sipior M.S., Fullton L.K., Zhang C.-Y., 
Schaefer K.G., 1999, AJ, 118, 488 

\noindent Corradi R.L.M., Ferrer O.E., Schwarz H.E., Brandi E., 
Garc\'{\i}a L., 1999, A\&A, 348, 978

\noindent Durand S., Acker A., Zijlstra A., 1998, A\&AS, 132, 13

\noindent Garc\'{\i}a-Segura G., 1997, ApJ, 489, L189

\noindent Garc\'{\i}a-Segura G., Langer N., R\'ozyczka M., Franco J., 1999, 
ApJ, 517, 767

\noindent Guerrero M.A., Manchado A., 1999, ApJ, 522, 378

\noindent Hollis J.M., Pedelty J.A., Lyon R.G., 1997, ApJ, 482, L85

\noindent Hollis J.M., Wagner R.M., Oliversen R.J., 1990, ApJ, 351, L17

\noindent Kwok S., Aaquist O.B., 1993, PASP, 105, 1456 (KA93)

\noindent Kostyakova E., 1992, PAZh, 18, 802

\noindent Livio M., Pringle J.E., 1996, ApJ, 465, L55

\noindent Mastrodemos N., Morris M., 1998, ApJ, 497, 303

\noindent Mastrodemos N., Morris M., 1999, ApJ, 523, 357

\noindent Mezger P.G., Henderson A.P., 1967, ApJ, 147, 490

\noindent Morris M., 1987, PASP, 99, 115

\noindent Miranda L.F., 1995, A\&A, 304, 531 (M95)

\noindent Miranda L.F., 1999, in ASP Conf. Ser., vol.188, p. 257

\noindent Miranda L.F., V\'azquez R., Corradi R.L.M., Guerrero M.A., 
L\'opez J.A., Torrelles J.M., 1999, ApJ, 520, 714

\noindent Reyes-Ruiz M., L\'opez J.A., 1999, ApJ, 524, 952

\noindent Richards A.M.S., Bode M.F., Eyres S.P.S., Kenny H.T., Davis R.J., 
Watson S.K., 1999, MNRAS, 305, 380

\noindent Sahai R., Dayal A., Watson A.M., et al., 1999, AJ, 118, 468

\noindent Sahai R., Trauger J.T., 1998, AJ, 116, 1357 

\noindent Schneider S.E., Terzian Y., Purgathofer A., Perinotto M., 1983, 
ApJS, 52, 399

\noindent Seaquist E.R., Taylor A.R., 1990, ApJ, 349, 313

\noindent Soker N., 1994, MNRAS, 270, 774

\noindent Soker N., 1996, ApJ, 468, 774

\noindent Soker N., 1998, ApJ, 496, 833

\noindent Soker N., 1999, AJ, 118, 2424

\noindent Soker N., Livio M., 1994, ApJ, 421, 219

\noindent Soker N., Rappaport S., Harpaz A., 1998, ApJ, 496, 842

\noindent Solf J., 1984, A\&A, 139, 296

\noindent Solf J., Ulrich , 1985, A\&A, 148, 274

\noindent Zhang C.Y., Kwok S., 1998, ApJS, 117, 341

\noindent V\'azquez R., L\'opez J.A., Miranda L.F., Torrelles J.M., 
Meaburn J., 1999, MNRAS, 308, 939

\noindent V\'azquez R., L\'opez-Mart\'{\i}n L., Miranda L.F., Esteban C., 
Torrelles J.M., Arias L., Raga A.C., 2000, A\&A, 357, 1031

\newpage

\noindent {\bf Figure captions}

\noindent {\bf Figure 1.} H$\alpha$, [OIII]5007 and [NII]6583 contour maps 
of Hu\,2-1 from the ground-based images. North is up, east to the left. The 
contours are logarithmic separated by a factor 1.78 in intensity and have 
been chosen in order to show up the main features of the nebula. The 
identified morphological components are indicated in the [NII] image 
(see text).

\noindent {\bf Figure 2.} HST image of Hu\,2-1 obtained in the [NII]6583 
line. Two different grey levels have been chosen in order to show both the 
faint and bright nebular structures. The morphological components are 
indicated (see also Fig.1).

\noindent {\bf Figure 3.} Grey-scale map of the central regions of Hu\,2-1 as 
observed in the H$\alpha$ line with the HST. The inner shell is indicated.

\noindent {\bf Figure 4.} Grey-scale ratio maps of Hu\,2-1 obtained from the 
images in Fig.1. North is up, east to the left. In each image ratio, black 
regions refer to high ratios, white regions refer to low ratios.

\noindent {\bf Figure 5.} Grey-scale representation of the 3.6\,cm continuum 
map with uniform weighting (beam 0$\rlas$20$\times$0$\rlas$19, 
PA $-$14$^{\circ}$). The grey levels (top scale) are in mJy\,beam$^{-1}$.

\noindent {\bf Figure 6.} Contour map of the velocity-integrated H92$\alpha$ 
emission (solid line) superposed on a contour map of the 3.6\,cm radio 
continuum emission (dashed line) made with natural weighting and a Gaussian 
taper of 500 k$\lambda$ which resulted in a synthesized beam of 
0.42$\times$0.40, PA 60$^{\circ}$. The levels for H92$\alpha$ are 30, 40, 
50, 60, 70, 80, 90 and 95 percent of the maximum of 50.2 
mJy\,beam$^{-1}$\,km\,s$^{-1}$. Radio continuum contour levels are 3, 5, 7, 
9, 15, 30, 50, 70, 90, 100, 150 and 200 $\times$ 7.1$\times$10$^{-2}$ 
mJy\,beam$^{-1}$, the rms noise of the map. The peak position of the radio 
continuum map is $\alpha$(1950) = 18$^h$ 47$^m$ 38${{\rlap.}^s}$60 and 
$\delta$(1950) = 20$^\circ$ 47$^\prime$ 08${\rlap.}^{\prime\prime}$0.

\noindent {\bf Figure 7.} H92$\alpha$ emission profile. The dotted line 
shows a Gaussian fit to the line profile.

\noindent {\bf Figure 8.} Contour maps of the H92$\alpha$ emission for the 
two velocity intervals indicated in the upper right corner. Solid contours 
represent blueshifted emission, dashed contours represent redshifted 
emission with respect to the systemic velocity.

\noindent {\bf Figure 9.} Position-velocity contour maps of the [NII]6583 
long-slit spectra. Position angle of the slit is indicated in the upper right 
corner. The contour are logarithmic separated a factor 2 in intensity. The 
knots C1, C2, C3 and C4 are indicated. The position of the central star is 
represented by the position of the intensity maximum of the continuum. 
V$_{\rm sys}$(shell) indicates the systemic velocity of the main nebular shell 
whereas V$_{\rm sys}$(C1--C2) and V$_{\rm sys}$(C3--C4) indicate the systemic 
velocity of the two pairs of compact knots (see text for details).

\vfil\eject

\begin{center}

{\bf Table\,1: Physical parameters of Hu\,2-1$^a$}

\begin{tabular}{lc}
\noalign{\smallskip}
\hline
\noalign{\smallskip}
Parameter                     &  Value   \\
\hline
\noalign{\smallskip}
S$_{\nu}$(continuum) (mJy)$^b$    & 110 $\pm$ 1  \\
S$_{\nu}$(line) (mJy)$^b$         & 3.9 $\pm$ 0.3  \\
V$_{LSR}$ (km\,s$^{-1}$)$^c$      & 37 $\pm$ 2 \\
$\Delta$V$_L$ (km\,s$^{-1}$)$^c$  & 33 $\pm$ 2 \\
T$_e$ (K)$^d$                     & 7000 $\pm$ 500 \\
N$_e$  (cm$^{-3}$)$^e$            & 5900 $\pm$ 100   \\
\noalign{\smallskip}
\hline
\end{tabular}

\end{center}

\vspace{0.3cm}

\noindent $^a$ Obtained from continuum and H92$\alpha$ observations at 
3.6\,cm following the formulation by Mezger \& Henderson (1967). \\
\noindent $^b$ Total continuum and integrated line flux density.\\
\noindent $^c$ Systemic velocity and velocity width of the H92$\alpha$ line, 
determined by a Gaussian fit. \\
\noindent $^c$ Electron temperature, assuming a distance of 2.35 kpc 
(see text). \\
\noindent $^e$ Electron density.\\

\vspace{2.0cm}

\begin{center}

{\bf Table\,2: Parameters of the knots C1, C2, C3 and C4$^a$}

\begin{tabular}{lccccc}
\noalign{\smallskip}
\hline
\noalign{\smallskip}
Knot & PA & V$_r$$^b$ & X$^c$  & $\Delta$V$^d$  & $\Delta$X$^e$ \\  
\noalign{\smallskip}
     & (degrees) & (km\,s$^{-1}$) & (arcsec) & (km\,s$^{-1}$) & (arcsec) \\
\hline
\noalign{\smallskip}  
C1   & 320 & $-$59  & 3.0  &  28 & 1.4 \\
C2   & 140 & $+$48  & 3.0  &  17 & 1.5 \\
C3   & 351 & $-$66  & 2.5  &  29 & 1.3 \\
C4   & 171 & $+$50  & 2.4  &  19 & 1.6 \\
\noalign{\smallskip}
\hline
\end{tabular}

\end{center}

\vspace{0.3cm}

\noindent$^a$ Obtained from the long-slit [NII] spectra shown in Fig.9. \\ 
$^b$ Radial velocity with respect to the systemic velocity of the main 
nebular shell (V$_{LSR}$ = 34.5 km\,s$^{-1}$).\\
$^c$ Angular distance from the central star along the corresponding PA.\\ 
$^d$ Velocity width (FWHM) corrected of spectral resolution.\\
$^e$ Spatial extent (FWHM) corrected of spatial resolution.\\

\end{document}